\documentclass[
showpacs,
floatfix,
aps,
prl,
amsmath,
twocolumn,
superscriptaddress,
tightenlines
groupaddress,
]{revtex4}

\usepackage{graphicx}
\usepackage{dcolumn}
\usepackage{amsmath}
\usepackage{amsfonts}
\usepackage{bm}
\usepackage{epsfig}
\usepackage{times}

\newcommand{\be}{\begin{equation}}
\newcommand{\ee}{\end{equation}}
\newcommand{\bea}{\begin{eqnarray}}
\newcommand{\eea}{\end{eqnarray}}

\renewcommand{\vec}[1]{{\bm #1}}
\newcommand{\an}{a^{(N)}_{\varphi\varphi'}}

\newcommand{\wc}{\omega_{\rm c}}

\newcommand{\Rc}{R_{\mbox{\scriptsize c}}}

\newcommand{\pF}{p_{\mathrm{F}}}
\newcommand{\Pmw}{{\cal P}_{\omega}}
\newcommand{\vare}{\varepsilon}

\def\Edc{\mathcal{E}_{\mathrm{dc}}}

\def\eac{\epsilon_{\mathrm{ac}}}
\def\edc{\epsilon_{\mathrm{dc}}}

\def\oc{\omega_{\mbox{\scriptsize {c}}}}

\def\rc{R_{\mbox{\scriptsize {c}}}}

\def\tauq{\tau_{\mbox{\scriptsize {q}}}}
\def\tautr{\tau_{\mbox{\scriptsize {tr}}}}

\newcommand{\req}[1]{Eq.~(\ref{#1})}

\bibpunct{[}{]}{,\!}{n}{,}{,}
\begin{document}

\title{
Non-linear Magnetoresistance Oscillations in Intensely Irradiated Two-Dimensional Electron Systems Induced by Multi-Photon Processes
}
\author{M.\,Khodas}
\affiliation{Department of Condensed Matter Physics and Materials Science, Brookhaven National Laboratory, Upton, NY 11973, USA}

\author{H.\,-S. Chiang}
\affiliation{School of Physics and Astronomy, University of Minnesota, Minneapolis, Minnesota 55455, USA}

\author{A.\,T. Hatke}
\affiliation{School of Physics and Astronomy, University of Minnesota, Minneapolis, Minnesota 55455, USA}

\author{M.\,A. Zudov}
\affiliation{School of Physics and Astronomy, University of Minnesota, Minneapolis, Minnesota 55455, USA}

\author{M.\,G. Vavilov}
\affiliation{Department of Physics, University of Wisconsin, Madison, WI 53706, USA }

\author{L.\,N. Pfeiffer}
\affiliation{Princeton University, Department of Electrical Engineering, Princeton, NJ 08544, USA}

\author{K.\,W. West}
\affiliation{Princeton University, Department of Electrical Engineering, Princeton, NJ 08544, USA}

\begin{abstract}
We report on magneto-oscillations in differential resistivity of a two-dimensional electron system subject to {\em intense} microwave radiation.
The period of these oscillations is determined not only by microwave frequency but also by its intensity.
A theoretical model based on quantum kinetics at high microwave power captures all important characteristics of this phenomenon which is strongly nonlinear in microwave intensity.
Our results demonstrate a crucial role of the multi-photon processes near the cyclotron resonance and its harmonics in the presence of strong dc electric field and offer a unique way to reliably determine the intensity of microwaves acting on electrons.
\end{abstract}
\maketitle

Over the past decade an array of remarkable effects was discovered in very high Landau levels of high mobility two-dimensional electron systems (2DES).
Among these are microwave-induced resistance oscillations (MIRO)~\citep{miro:exp}, phonon-induced resistance oscillations~\citep{piro:exp}, Hall field-induced resistance oscillations~\citep{hiro:exp,hatke:2009c}, zero-resistance states~\citep{zrs}, and zero-differential resistance states~\citep{zdrs}.
Theories of magneto-resistance oscillations are based on the quantum kinetic description and consider two main mechanisms: 1) the ``displacement'' mechanism originating from modification of impurity scattering by microwave (ac) or dc electric fields~\cite{disp:th,vavilov:2004,vavilov:2007} and 2) the ``inelastic'' mechanism stepping from the formation of the non-equilibrium energy distribution~\citep{inel:th}.
MIRO are controlled by the ratio of microwave frequency $\omega=2\pi f$ to the cyclotron frequency $\oc=eB/m^*$, $\eac\equiv\omega/\oc$.
Hall field induced oscillations, which appear in differential resistivity, are governed by $\edc\equiv e\Edc(2\rc)/\hbar\oc$, where $\Edc$ is the Hall field, $2\rc=2v_F/\wc$ is the cyclotron diameter, and $v_F$ is the Fermi velocity.
Finally, in a 2DES subject to both ac and dc fields, the resulting oscillations were found to depend on simple combinations of ac and dc parameters, i.e. $\eac\pm\edc$~\citep{hatke:2008,khodas:2008}.
Such dependence indicates that the relevant inter-Landau level scattering processes involved a {\em single} photon.

Importance of processes involving {\em multiple} microwave quanta was suggested by numerous experiments~\citep{fmiro:exp} reporting MIRO-like features in the vicinity of {\em fractional} values of $\eac$.
The most prominent series of, so called, fractional MIRO occurs near subharmonics of the cyclotron resonance, $\eac=1/2, 1/3, 1/4, ...$ .
Theoretical proposals considered both {\em multi-photon}~\citep{lei,dmitriev:2007b} and {\em single-photon}~\citep{pechenezhskii:2007,dmitriev:2007b} mechanisms to explain the response at fractional $\eac$.
On the other hand, relevance of multi-photon processes near the cyclotron resonance and its harmonics was not considered either experimentally or theoretically.

In this paper we report on another class of magnetoresistance oscillations in a high-mobility 2DES exposed to \emph{high-power} microwave radiation and \emph{strong} dc electric field.
These oscillations are manifested by a series of \emph{multiple} maxima and minima all occurring in the close proximity to the cyclotron resonance and its \emph{harmonics}.
Furthermore, the period and the phase of these oscillations depend not only on $\eac$ and $\edc$ \emph{but also} on the microwave intensity.
This characteristic sensitivity to microwave intensity sets apart this strongly non-linear phenomenon from all previously reported resistance oscillations.

To explain our experimental findings we propose a theoretical model based on quantum kinetics which captures all important characteristics of the phenomenon.
As we will show, this unusual effect owes to the quantum oscillations in the density of states and a crucial role played by \emph{multi-photon} processes.
In the presence of radiation the electron states are split into Floquet subbands separated by $\hbar \omega$ in a similar way as quasi-particle states split in the course of \emph{photon-assisted} tunneling across a Josephson junction~\cite{Tien:1963}.
The scattering rate off disorder is then controlled by the overlap of such subbands leading to oscillatory  behavior in differential magnetoresistance.

Our experiment was performed on a 100 $\mu$m-wide Hall bar etched
from a symmetrically doped GaAs/AlGaAs quantum well. After brief
low-temperature illumination with visible light, density and
mobility were $n_e\simeq 3.8\times 10^{11}$ cm$^{-2}$ and $\mu\simeq
1.3\times 10^{7}$ cm$^{2}$/Vs, respectively. All the data were
recorded under continuous irradiation by $f=27$ GHz microwaves in a
Faraday geometry at $T\simeq$ 1.5 K. Differential resistivity,
$r(I)\equiv dV/dI$, was measured using a quasi-dc (a few Hertz)
lock-in technique.

\begin{figure}[t]
\includegraphics{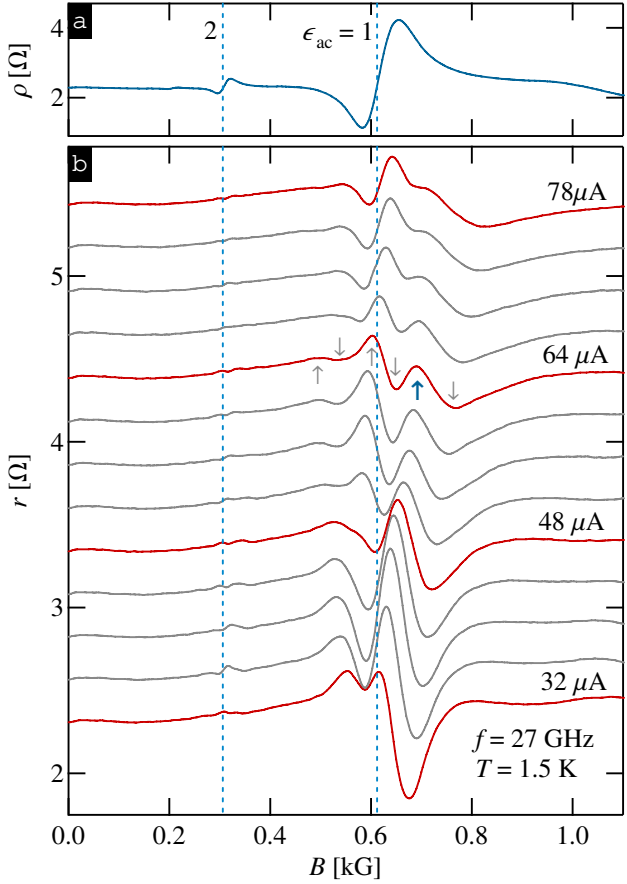}
\vspace{-0.1in}
\caption{[color online]
(a) Magnetoresistivity $\rho(B)$.
(b) Differential resistivity $r(B)$ measured at dc currents from $I = 32$ to $78$ $\mu$A in step of 4 $\mu$A.
All data are acquired under irradiation by $f=27$ GHz microwaves at $T=1.5$ K.
Vertical lines mark positions of the cyclotron resonance and its second harmonic.
}
\vspace{-0.1in}
\label{data}
\end{figure}
As shown in Fig.\,\ref{data}\,(a), the magnetoresistivity measured under microwave irradiation in the absence of dc field exhibits sharp MIRO features around the cyclotron resonance ($\eac=1$) and its second harmonic ($\eac=2$).
In both cases we observe exactly {\em one} minimum and {\em one} maximum positioned roughly symmetrically about $\eac=1,2$ (cf.\, vertical lines).

In Fig.\,\ref{data}\,(b) we present differential magnetoresistivity $r(B)$ measured at dc currents $I$ from 32 $\mu$A to 78 $\mu$A (in step of 4 $\mu$A) at the same radiation frequency and intensity.
Remarkably, the data readily reveal {\em multiple} maxima and minima occurring in the proximity to the cyclotron resonance.
For example, the data obtained at $I=64$ $\mu$A exhibit three maxima (cf.,\,$\uparrow$) and three minima (cf.,\,$\downarrow$).
With increasing $I$ all oscillations move to higher magnetic fields in a regular fashion.
While not so pronounced, similar behavior is observed near $\eac=2$.
These findings indicate that these magneto-oscillations are qualitatively different from all reported previously.
As we show below, the phenomenon owes to a non-trivial role played by multi-photon processes at integer $\eac$ in the regime of strong microwave and dc electric fields.

To interpret the experimental data, the dynamical screening of microwaves by 2DES~\citep{pechenezhskii:2007,quinn:1976} has to be considered.
Our theoretical results are expressed through the dimensionless microwave power $\Pmw$ which includes this effect explicitly,
\begin{equation}\label{Pmw+}
\Pmw^{\pm} = \frac{ {\mathcal S} }{\epsilon_{0} c \hbar^{2} }\frac{ (2 e R_c)^2 }{ \tau_{ \mathrm{em}  }^{-2} +(\omega\pm\wc)^2} \, .
\end{equation}
Here, ${\mathcal S }=\epsilon_{0} c {\mathcal{E}}_{\mathrm{ac}}^2$ is the energy flux carried by the microwave radiation, $ \tau_{ \mathrm{em}  }= \epsilon_{0}(\sqrt{ \epsilon }   + 1 ) m^* c /2 n e^{2}$ is the electromagnetic damping time, and $\epsilon = 12.8$ is the dielectric constant of GaAs. 
In our 2DES, $   \tau_{ \mathrm{em}  }  \simeq 3.8$ ps and $\oc\tau_{\mathrm{em}}\sim 1$. 
For that reason the parameter $\Pmw^{\pm}$ has a weak magnetic field dependence near the classical resonance $\wc = 2\pi f$, which is the focus of the present study. 
At the same time, we can still retain only the active component $\Pmw^-$ of the incoming radiation (below the ``$-$'' sign is omitted). 
At $\omega  \approx \omega_c$, $\Pmw=1$ corresponds to a microwave flux of one photon passing through the area of $\lambda_{H}^{2} = \hbar c/ e B$ during the time $\tau_{ \mathrm{em}  }$.

We now present the main result of our calculations which is the expression for the differential resistivity
$\delta r = r - r_D$:
\begin{equation} \begin{split}\label{Result}
&\frac{\delta r }{ r_D }\!=\!
\frac{ (4 \lambda)^2 \tautr }{ \pi \tau_{\pi} } \Big[
\cos 2 \pi \edc
J_0\left(4 \sqrt{\Pmw } \sin \pi \eac \right) \\
 &\,-\!\frac{ 2 \eac }{ \edc }\!\sqrt{\Pmw } \sin 2 \pi \edc \cos \pi \eac
J_1\left(4 \sqrt{\Pmw } \sin \pi \eac \right) \Big] .
\end{split}
\end{equation}
Here, $J_n(x)$ is the Bessel function of $n$-th order, $\tau_{\rm tr}$ and $\tau_{\pi}$ are transport scattering and backscattering times, $r_D$ is the Drude resistivity, and $\lambda=\exp[-\pi/\wc\tauq]$ is the Dingle factor.
In our 2DES, the quantum scattering time $\tauq \simeq 20$ ps, which corresponds to $\lambda^2\simeq 0.15$ at the cyclotron resonance.
Equation (\ref{Result}) holds at high temperatures, $T\gtrsim \hbar \omega\sqrt{\Pmw}$ and strong direct currents, $\edc\gtrsim \sqrt{\Pmw}$.

We now discuss the evolution of the oscillation maxima positions with dc field.
At $\Pmw\gtrsim 1$, Eq.~\eqref{Result} reduces to
\begin{equation*}
\frac{ \delta r }{ r_D }\nonumber
\propto
\cos\!\left[ 4 \sqrt{\Pmw} \sin( \pi \eac ) - \pi/4 - 2 \pi \edc \right] .
\end{equation*}
Thus the position of the maximum closest to the cyclotron resonance is given by $\eac^+=1 + \delta$, where $\delta$
depends linearly on $\edc$:
\begin{align} \label{delta}
\delta(\edc) = - \frac 1 { 2 \sqrt{\Pmw} } \left ( \edc + \frac 1 8 \right ).
\end{align}
This expression allows us to extract the value of $\Pmw$ which cannot be reliably determined in conventional, e.g. MIRO, experiments.
\begin{figure}[t]
\includegraphics{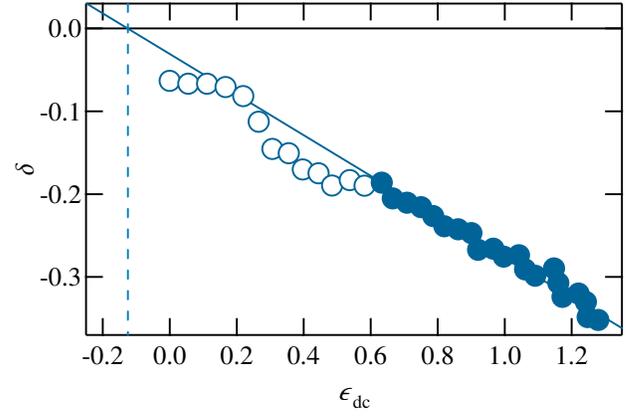}
\vspace{-0.1in}
\caption{[color online] Offset of the resistivity peak near the cyclotron resonance, $\delta = \eac^+ -1$ (open circles) vs. $\edc$.
Linear fit (solid line) performed for $\edc > 0.6$ (solid circles) crosses $\delta =0$ at $\edc = - 1/8$ (dashed line) as prescribed by Eq.~\eqref{delta}.
}
\vspace{-0.1in}
\label{fig2}
\end{figure}
In Fig.~\ref{fig2} we present experimental evolution of the phase $\delta = \eac^+ - 1$ with increasing $\edc$ and observe linear dependence for $\edc > 0.6$.
The linear fit (solid line) generates $\Pmw = 4.2$ and passes through $(\delta,\edc)=(0,-1/8)$, as prescribed by Eq.~\eqref{delta}. %

\begin{figure}[t]
\includegraphics{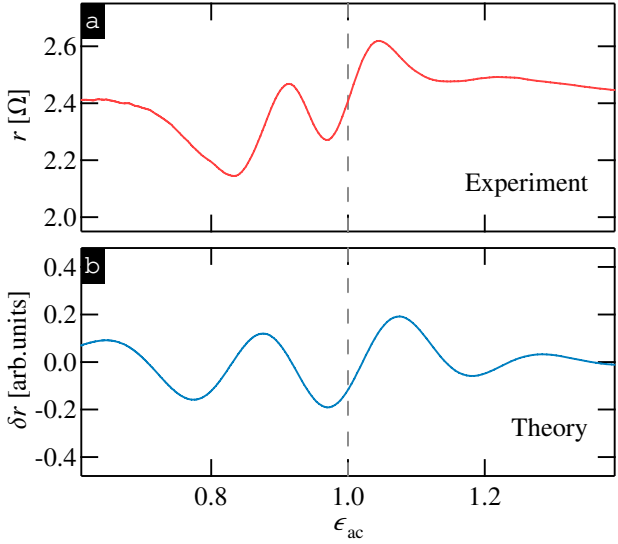}
\vspace{-0.1in}
\caption{[color online]
(a) Differential resistivity $r$ vs. $\eac$ measured at dc current $I = 54$ $\mu$A under irradiation by microwaves.
(b) Correction to the differential resistivity calculated with Eq.\,(\ref{Result}) using $ \Pmw =4.2$ and $I = 54$ $\mu$A.
Vertical line marks position of the cyclotron resonance.
}
\vspace{-0.1in}
\label{fig3}
\end{figure}

Having extracted $\Pmw$ and obtained excellent agreement for the evolution of the peak position with $\edc$, we can now compare the experimental and theoretical curves directly.
Fig.\,\ref{fig3}(a) shows our data obtained in microwave-irradiated 2DES carrying direct current $I = 54$ $\mu$A, which corresponds to $\edc\simeq 1.65$ at $\eac=1$.
In Fig.\,\ref{fig3}(b) we present the results of calculations using \req{Result} with $\Pmw=4.2$ extracted earlier.
We observe that the theoretical curve recreates experimental data, including the number of measured oscillations, reasonably well.
The theory also captures the amplitude of the observed oscillations.
Using \req{Result} with $\tautr/\tau_\pi\approx 0.18$ measured in recent non-linear dc transport experiments~\citep{hatke:2009c} we arrive at $\delta r/r_D \simeq 0.1$.
This value agrees well with the experimental data in Fig.~\ref{fig3}(a) and further reinforces the validity of our model as well as the procedure for extracting~$\Pmw$.

The remaining part of the paper is devoted to the derivation of the main theoretical result, \req{Result}.
Our derivation is based on the quantum kinetic equation of electrons in the presence of in-plane electric fields in the limit of overlapping Landau levels.
We find the distribution function
$f_{\varepsilon;\varphi}$ which depends on the direction of the momentum $\vec{p} = p_{\mathrm{F}} \vec{n}_{\varphi}$ of an electron at the Fermi surface in the quasi-classical approximation, $E_F \gg \omega_c$.
The current density is given by $j = 2 e v_{F} \int \cos \varphi \nu(\epsilon) f_{\varepsilon;\varphi} d \epsilon d \varphi / 2 \pi $.

The distribution function satisfies the kinetic equation~\citep{vavilov:2004,khodas:2008}
\begin{equation} \label{kin-eq}
\omega_c \partial_{\varphi } f_{\varepsilon;\varphi} = \mathrm
{St}\{ f \}_{\varepsilon;\varphi} .
\end{equation}
with the collision term given in the Floquet representation~\citep{moskalets:2004} as
\begin{equation} \label{collision}
\mathrm {St} \{ f \}\!=\!
\sum_N\!\! \int\!\!\frac{ d \varphi'}{ 2 \pi }
\Gamma^{(\!N\!)}_{ \varphi \varphi' } \!\!\left[ f_{\varphi'}( \varepsilon\!+\!W_{\varphi\varphi'}\! +\! N \hbar \omega) \!-\!
f_{\varphi}( \varepsilon ) \right] .
\end{equation}
Equation \eqref{collision} accounts for scattering events whereby the particle changes its momentum by
 $ \Delta \vec{q} = \pF \left( \vec{n}_{\varphi}- \vec{n}_{\varphi'} \right)$ and absorbs ($ N > 0$) or emits ($ N < 0$) $|N|$ photons.
The kinetic energy change includes the work done by the dc electric field as the result of
shift of the guiding center, $W_{\varphi\varphi'} = e E \Rc (\sin \varphi - \sin \varphi' ) $,
and the energy of absorbed or emitted photons. The scattering rate is
\begin{equation} \label{rate}
\Gamma^{(\!N\!)}_{ \varphi \varphi' } = \frac{ \an }{ \tau_{\varphi\varphi'} } \nu_{0}^{-1} \nu(\varepsilon\!+\!W_{\varphi\varphi'}\! +\! N \hbar\omega)\, ,
\end{equation}
where
\begin{equation}
\label{eq:an}
\an\!=\!\!\!\int\limits_{0}^{2\pi/ \omega }\!\!\!\!\frac{\omega d t }{2\pi}e^{i N \omega t} J_0 \left( 4 \sqrt{\Pmw }
\left| \sin \frac{\varphi-\varphi'}{2} \right|\sin \frac{\omega t}{ 2 }\right)
,
\end{equation}
the density of states is
$\nu(\epsilon) = \nu_0\left[ 1 - 2 \lambda \cos( 2 \pi \epsilon/\hbar \wc ) \right]$ with
$\nu_0 = m / \hbar^{2} \pi$ being the DoS per spin.
In the classical limit of constant DoS the normalization condition
$\sum_{N} \Gamma^{(\!N\!)}_{ \varphi \varphi' } = 1/\tau_{ \varphi \varphi' } $ holds, as the radiation changes electron energy distribution while not affecting the total number of particles scattered into a given angular range.
We look for the solution of Eq.~\eqref{kin-eq} in the form $f = f_{T} + \Delta f_{\rm cl} + \lambda \Delta^{(1)} f$, where $f_{T}$ is the
Fermi-Dirac distribution function, $f_{T}+\Delta f_{\rm cl}$ is the solution of the classical kinetic equation
with $\lambda=0$, and
\begin{equation} \label{correction}
\Delta^{(1)} f(\vare,\varphi) =
A_1 \cos \varphi \frac{\partial f_T(\vare)}{\partial \vare} \cos \frac{ 2 \pi \vare}{ \hbar \omega_c } \, .
\end{equation}

We substitute Eq.~\eqref{correction} to Eqs.~\eqref{kin-eq}, \eqref{collision} and \eqref{rate} and neglect terms containing derivatives of the Fermi distribution of the second and higher orders since the contributions of such terms vanish upon the integration over the energy.
We obtain
\begin{eqnarray}\label{A1}
A_1 & = & \frac{ 4 }{\wc }\left\langle \sin \varphi \sum_N
K^{(N)}_{\varphi\varphi'} \right\rangle_{\varphi\varphi'}\, , \\
K^{(N)}_{\varphi\varphi'} & = & \frac{ \an }{ \tau_{\varphi\varphi'} } \left( W_{\varphi\varphi' } + N \hbar \omega \right)
\cos \left[ \frac{2 \pi \left(W_{\varphi \varphi'} + N \hbar \omega \right) }{ \hbar \omega_c } \right]
\, . \notag
\end{eqnarray}
The angular integrations in Eq.~\eqref{A1} can be carried out using
the stationary phase approximation. At $\edc \gtrsim
\mathrm{max}\{\sqrt{\Pmw}, 1\}$ the main contribution to the
integral comes from narrow intervals centered at $\bar{\varphi}= \pm
\pi/2$, $\bar{\varphi}'= \mp \pi/2$. This corresponds to
backscattering of electrons in strong dc electric field. To
calculate $A_1$, we substitute $a^{N}_{\bar{\varphi}\bar{\varphi}'
}$ from \req{eq:an} to \req{A1} and sum over $N$. The summation
results in a set of $\delta$-functions $\delta(\omega t\pm 2\pi
\omega/\wc+2\pi k)$ and their derivatives ($k=0,\pm 1,\pm 2,\dots$).
The remaining time integration reproduces the final result of this
calculation presented in \req{Result}.

Without microwave radiation
or in the weak power limit, \req{Result} reduces to the result of
Refs.~\citep{vavilov:2007} or~\citep{khodas:2008}, respectively.
At stronger microwave powers,
the wave functions are of the Floquet type\cite{moskalets:2004} and characterized by an additional time-dependent phase factor.
According to Eq.~\eqref{eq:an},
the number of photons maximizing $a^{N_m}_{\bar{\varphi}\bar{\varphi}' }$ is $N_{m} = \sqrt{\Pmw}$, since
in this case the phase factor oscillates in-phase with factor $\exp(iN\omega t)$ and the most efficient scattering off disorder is accompanied by absorption or emission of $N_{m}$ quanta of energy $\hbar\omega$,  Eq.~\eqref{rate}.

In summary, we reported on experimental and theoretical studies of magnetotransport properties of high-mobility two-dimensional electron systems driven away from equilibrium by {\em intense} microwave radiation and {\em strong} dc electric fields.
We observed a new class of magnetoresistance oscillations near the cyclotron resonance and its harmonics,
with positions of peaks and dips strongly dependent on microwave intensity.
We proposed a theoretical model based on the quantum kinetic equation taking full account of multi-photon processes.
Our calculations capture all important characteristics of the phenomenon $-$ the period, the phase, and the amplitude of the oscillations are all in excellent agreement with experimental observations.
Taken together, these results demonstrate the crucial role of the multi-photon processes near the cyclotron resonance and its harmonics in the presence of strong dc electric field and offer a unique way to reliably determine the microwave intensity seen by 2D electrons.

We thank I. A. Dmitriev and B. I. Shklovskii for useful discussions and remarks.
The work at Minnesota was supported by NSF Grant No. DMR-0548014.
M.K. acknowledges support by the US DOE under contract number DE-AC02-98 CH 10886 and BNL grant 08-002.
M.V. acknowledges the Donors of the American Chemical Society Petroleum Research Fund for partial support of this research.

\end{document}